\documentclass[a4paper,11pt]{article}
\usepackage{amssymb,amsmath,amsthm,epsfig,psfrag}

\usepackage{mathrsfs}

\newcommand{\RR}{{\mathbb{R}}}

\newcommand{\ZZ}{{\mathbb{Z}}}

\newcommand{\pa}{\partial}
\newcommand{\ii}{{\rm i}}
\newcommand{\dd}{{\rm d}}

\newcommand{\sfrac}[2]{{\textstyle\frac{#1}{#2}}}
\newcommand{\Tr}{\mathrm{Tr}}
\newcommand{\curlyv}{{\mathscr V}}

\theoremstyle{plain}
\newtheorem{theorem}{Theorem}
\newtheorem{lemma}[theorem]{Lemma}
\newtheorem{proposition}[theorem]{Proposition}

\theoremstyle{definition}
\newtheorem{definition}[theorem]{Definition}

\title{On the charge density \\ and asymptotic tail of a monopole}
\author{Derek Harland${}^\ast$ and Daniel Nogradi${}^\dagger$
  \bigskip
  \\${}^\ast$ School of Mathematics, University of Leeds, Leeds, UK
  \medskip
  \\d.g.harland@leeds.ac.uk
  \bigskip
  \\${}^\dagger$ Eotvos University, Institute for Theoretical Physics, \\Budapest 1117, Hungary
  \medskip
  \\MTA-ELTE Lendulet Lattice Gauge Theory Research Group, \\1117 Budapest, Hungary
  \medskip
  \\nogradi@bodri.elte.hu}
\date{13th August 2015}

\begin{document}

\maketitle

\begin{abstract}
We propose a new definition for the abelian magnetic charge density of a non-abelian monopole, based on zero-modes of an associated Dirac operator.  Unlike the standard definition of the charge density, this density is smooth in the core of the monopole.  We show that this charge density induces a magnetic field whose expansion in powers of $1/r$ agrees with that of the conventional asymptotic magnetic field to all orders.  We also show that the asymptotic field can be easily calculated from the spectral curve. Explicit examples are given for known monopole solutions.
\end{abstract}

\section{Introduction}

Non-abelian monopoles are smooth, static, finite-energy solutions to the Yang-Mills-Higgs equations with non-abelian gauge group.  It was first noticed by 't Hooft and Polyakov that to a distant observer they resemble Dirac monopoles in an abelian gauge theory \cite{tHooft74, polyakov74}.  Thus the singularity of the Dirac monopole can be smoothed out in non-abelian gauge theory.

't Hooft defined \cite{tHooft74} an asymptotic abelian magnetic field of a non-abelian monopole with gauge group SU(2).  The flux of this magnetic field through the two-sphere at infinity is topologically quantised and non-zero.  However, in the core of the monopole this magnetic field is singular \cite{prasad-sommerfield75}, and the magnetic charge distribution which induces it typically has delta-function singularities.  Thus, while a non-abelian monopole is smooth, the magnetic charge distribution associated to 't Hooft's magnetic field is far from being smooth.

As has been argued by Coleman \cite{coleman}, there is no reason to expect the abelian magnetic field of a non-abelian monopole to be uniquely defined: any magnetic field which agrees with 't Hooft's asymptotically is an equally viable candidate.  However, to date no definition of a magnetic field has been proposed which smoothes out the singularities in 't Hooft's field.  In this article we remedy this situation: we propose a novel definition of the magnetic charge density of a non-abelian monopole which, unlike 't Hooft's charge density, is smooth.  Moreover, we show that the magnetic field induced by this charge density agrees with 't Hooft's asymptotically, at least in the case of BPS monopoles.  Our charge density is evaluated by summing the squared norms of zero-modes of a Dirac operator.  In this way it resembles the trace of the Bergman kernel used in K\"ahler geometry, which is a sum of squared norms of zero-modes of a Cauchy-Riemann operator.

The proof that our charge density induces the correct asymptotic magnetic field is based on much of the mathematical formalism that has been developed to study BPS monopoles, including the Nahm transform and spectral curves.  A prominent role is played by a function which we call the \emph{tail} of the monopole.  This function describes the asymptotics of the Higgs field and was first studied by Hurtubise \cite{hurtubise}.  We have observed that this tail function can in many cases be calculated explicitly, which is remarkable given the paucity of explicit monopole solutions.  Another interesting consequence of our work is a proof of a conjecture \cite{BNvB, nogradi} relating conserved charges of the Nahm equation to asymptotics of an associated Greens' function.

This paper is structured as follows.  In section \ref{sec2} we review standard results relating moments of electric (or magnetic) charge distributions to the asymptotics of the electric (or magnetic) fields that they induce.  In section \ref{sec3} we introduce our charge density and show that its integral agrees with the magnetic flux through the two-sphere at infinity.  The proof that the magnetic field induced by this charge density agrees asymptotically with 't Hooft's proceeds in two parts: in section \ref{sec4} we show that the moments of the charge density equal certain conserved quantities of the Nahm equation, and section \ref{sec5} we show using Hurtubise' work on the tail function that these conserved quantities also prescribe the asymptotic expansion of 't Hooft's magnetic field.  In section \ref{sec6} we present some explicit calculations of the tail function.  We discuss promising extensions of this work in section \ref{sec7}.

\section{Moments and charge distributions}
\label{sec2}

It is well-known that the moments of a distribution of electric charge determine the asymptotic expansion of the induced electric field \cite{jackson}.  More precisely, let $\rho:\RR^3\to\RR$ be a smooth function that decays exponentially as $r\to\infty$ and let $\phi:\RR^3\to\RR$ be a potential of the induced electric field $e_i=-\pa_i\phi$.  The functions $\rho$ and $\phi$ are related by 
\[ \rho=\pa_ie_i=-\triangle\phi.\]
Suppose that $\phi$ has an expansion in powers of $1/r$ of the form
\begin{equation}
\label{expansion phi}
 \phi = \sum_{\ell=0}^{\infty} \frac{\phi_\ell(\theta,\varphi)}{r^{l+1}}.
\end{equation}
The functions $\phi_\ell$ must then be spherical harmonics of weight $\ell$, as $\triangle\phi$ decays exponentially.  The expansion \eqref{expansion phi} is called the multipole expansion of $\phi$.

Write
\[ x(\zeta)=\frac12(x_1+\ii x_2) + \zeta x_3 - \frac12(x_1-\ii x_2)\zeta^2, \]
and for each $\ell=0,1,2,\ldots$ let $Q_\ell(\zeta)$ be the polynomial
\begin{equation}
\label{moment}
 Q_\ell(\zeta) = \int_{\RR^3} \rho(\mathbf{x})x(\zeta)^\ell \dd^3 x.
\end{equation}
The $2l+1$ coefficients of $Q_\ell(\zeta)$ are moments of the distribution $\rho$.  They determine, and are determined by, the spherical harmonics $\phi_\ell$.  More precisely:
\begin{proposition}
\label{prop:moments asymptotics}
Let $\rho$ be an exponentially decaying function and let $\phi$ solve $\triangle\phi=-\rho$, such that $\phi$ has an expansion in powers of $1/r$ of the form \eqref{expansion phi}.  Then the coefficients of this expansion and the moments \eqref{moment} of $\rho$ satisfy the identities
\begin{align}
\label{phi to M}
 Q_\ell(\zeta)&=(2\ell+1) \int_{S^2} \phi_\ell(\theta,\phi)n(\zeta)^\ell\,\sin\theta\dd\theta\dd\varphi \quad\mbox{and} \\
\label{M to phi}
 \phi_\ell(\theta,\varphi)&=\frac{1}{8\pi^2\ii}\oint_{\Gamma} \frac{Q_{\ell}(\zeta)}{n(\zeta)^{\ell+1}}\dd\zeta,
\end{align}
in which
\[ n(\zeta):=\frac{x(\zeta)}{r} = \frac12 e^{\ii\varphi}\sin\theta + \zeta\cos\theta - \frac12\zeta^2 e^{-\ii\varphi}\sin\theta  \]
is a polynomial in $\zeta$ whose coefficients are spherical functions, and $\Gamma$ is a small contour which circles the point
\[ \zeta=-\frac{x_2+\ii x_3}{x_1+r}=-\frac{e^{\ii\varphi}\sin\theta }{\cos\theta+1}. \]
\end{proposition}

Although the equivalence of the moments and the spherical harmonics $\phi_\ell$ is a standard result, we present a brief proof of equations \eqref{phi to M} and \eqref{M to phi}, as our notation (in particular our choice of parameterising the moments using a polynomial) is non-standard.

\begin{proof}
The proof of \eqref{phi to M} rests on the fact that the coefficents of $x(\zeta)^l$ solve the Laplace equation.  This fact follows by induction from the following two identities, which are easily verified:
\begin{align*}
 \pa_i\pa_i x(\zeta) &= 0 \\
 \pa_i x(\zeta)\pa_i x(\zeta) &= 0.
\end{align*}
Integrating the right hand side of \eqref{moment} by parts twice and substituting the series expansion for $\phi$ then yields
\begin{align*}
 Q_\ell(\zeta) &= -\lim_{R\to\infty}\int_{\|\mathbf{x}\|\leq R} \triangle\phi\, x(\zeta)^\ell\,\dd^3x \\
 &= \lim_{R\to\infty} \int_{S^2_R} \left( \mathcal{\phi}\frac{\pa(x(\zeta)^\ell)}{\pa r} - x(\zeta)^\ell\frac{\pa\mathcal{\phi}}{\pa r} \right)  r^2\sin\theta\dd\theta\dd\varphi \\
 &= \lim_{R\to\infty} \sum_{m=0}^\infty R^{\ell-m} (\ell+m+1) \int_{S^2} \phi_m \,n(\zeta)^\ell\, \sin\theta\dd\theta\dd\varphi.
\end{align*}
Since $\triangle (x(\zeta)^\ell)=0$, the coefficients of $n(\zeta)^\ell$ are spherical harmonics of weight $\ell$.  Now spherical harmonics of different weights are $L^2$-orthogonal, so the integral over $S^2$ appearing in the preceding expression vanishes unless $\ell=m$.  Thus the expression reduces to the stated result \eqref{phi to M}.

The second identity \eqref{M to phi} follows from the first via representation theory.  The space of spherical harmonics of weight $\ell$ and the space of degree $2\ell$ polynomials both carry representations of $\mathfrak{su}(2)$: it is easily checked that the operators
\begin{align*}
L_{\pm} &= (n_2\pm\ii n_3)\frac{\pa}{\pa n_1}-n_1\left(\frac{\pa}{\pa n_2} \pm \ii\frac{\pa}{\pa n_3}\right), \\
L_0 &= n_2\frac{\pa}{\pa n_3}-n_3\frac{\pa}{\pa n_2} \\
J_+^{(\ell)} &= \zeta^2\frac{\pa}{\pa \zeta} -2\ell\zeta \\
J_-^{(\ell)} &= \frac{\pa}{\pa \zeta} \\
J_0^{(\ell)} &= \ii\left(\zeta\frac{\pa}{\pa\zeta}-\ell\right)
\end{align*}
obey the $\mathfrak{su}(2)$ commutation relations,
\begin{align*}
[L_0,L_\pm]&=\pm\ii L_\pm & [L_+,L_-]&=2\ii L_0 \\
[J^{(\ell)}_0,J^{(\ell)}_\pm]&=\pm\ii J^{(\ell)}_\pm & [J^{(\ell)}_+,J^{(\ell)}_-]&=2\ii J^{(\ell)}_0.
\end{align*}
These representations are all irreducible.

Equations \eqref{phi to M} and \eqref{M to phi} define maps between the spaces of spherical harmonics of weight $\ell$ and the space of degree $2\ell$ polynomials, and we aim to show that these maps are inverse to each other.  The maps respect the action of $\mathfrak{su}(2)$, in the sense that
\begin{align}
\label{J and L 1}
J^{(\ell)}_\mu \int_{S^2} \phi_\ell(\theta,\phi)n(\zeta)^\ell\,\sin\theta\dd\theta\dd\varphi &=\int_{S^2} L_\mu\phi_\ell(\theta,\phi)n(\zeta)^\ell\,\sin\theta\dd\theta\dd\varphi \\
\label{J and L 2}
L_\mu \frac{Q_{\ell}(\zeta)}{n(\zeta)^{\ell+1}}\dd\zeta &=\oint_{\Gamma} \frac{J^{(\ell)}_\mu Q_{\ell}(\zeta)}{n(\zeta)^{\ell+1}}\dd\zeta\qquad\mbox{for }\mu=0,\pm.
\end{align}
Their composition is a linear map from the space of spherical harmonics of weight $\ell$ to itself that commutes with the action of $\mathfrak{su}(2)$.  By Schur's lemma, this map is equivalent to multiplication by a constant, so we only need to show that this constant is 1.  We can do so by showing that the map fixes just one element.

We choose the element $\phi_\ell=(n_2-\ii n_3)^\ell$.  According to eq.\ \eqref{phi to M} the associated polynomial is
\begin{align*}
Q_\ell(\zeta) &= (2\ell+1) \int_{S^2} (n_2-\ii n_3)^\ell n(\zeta)^\ell\,\sin\theta\dd\theta\dd\phi \\
&= (2\ell+1) \int_{S^2} (n_2-\ii n_3)^\ell \left(\frac{n_2+\ii n_3}{2}\right)^\ell\,\sin\theta\dd\theta\dd\phi \\
&= \frac{(2\ell+1)(2\pi)}{2^{\ell}} \int_{0}^\pi \sin^{2\ell+1}\theta\,\dd\theta \\
&= \frac{2^\ell (\ell!)^2(4\pi)}{(2\ell)!}.
\end{align*}
We now evaluate the right hand side of eq.\ \eqref{M to phi} with this particular $Q_\ell$.  In order to evaluate the contour integral we factorise the denominator: we find that
\[ n(\zeta) = \frac{(\zeta-\zeta_-)(\zeta-\zeta_+)}{\zeta_+-\zeta_-} ,\mbox{ where } \zeta_{\pm} = -\frac{n_2+\ii n_3}{n_1\pm 1}, \]
and thus that
\begin{align*} (n(\zeta))^{-\ell-1} &= \frac{1}{(\zeta-\zeta_+)^{\ell+1}} \left(1-\frac{\zeta-\zeta_+}{\zeta_--\zeta_+}\right)^{-\ell-1}\\
&=\sum_{m=0}^\infty \left(\begin{array}{c}\ell+m \\ \ell \end{array}\right)\frac{(\zeta-\zeta_+)^{m-\ell-1}}{(\zeta_--\zeta_+)^m}.
\end{align*}
Only the $m=\ell$ term in this Laurent series contributes to the integral \eqref{M to phi}, so
\begin{align}
\frac{1}{8\pi^2\ii} \oint_{\Gamma} \frac{Q_\ell(\zeta)}{n(\zeta)^{\ell+1}} \dd\zeta 
&= \frac{2^\ell (\ell!)^2}{(2\ell)!}\frac{1}{2\pi\ii} \oint_{\Gamma} \left(\begin{array}{cc}2\ell \\ \ell\end{array}\right) \frac{1}{(\zeta_--\zeta_+)^\ell} \frac{1}{\zeta-\zeta_+}  \dd\zeta \nonumber \\
&= 2^\ell(\zeta_--\zeta_+)^{-\ell} \nonumber \\
\label{oint 1/n^l}
&= (n_2-\ii n_3)^\ell.
\end{align}
This equals the original function $\phi_\ell$, so eq.\ \eqref{M to phi} follows from eq.\ \eqref{phi to M} as claimed.
\end{proof}

\section{The charge density of a monopole}
\label{sec3}

The Yang-Mills Higgs energy for an $\mathfrak{su}(2)$ gauge field $A_i$ and adjoint scalar $\Phi$ on Euclidean $\RR^3$ is
\[ E = \int_{\RR^3}\left[-\frac{1}{4}\Tr\left( D_i\Phi D_i\Phi \right) - \frac{1}{8}\Tr\left( F_{ij}F_{ij}\right)+\lambda (1-\|\Phi\|^2)^2 \right] \dd^3x, \]
in which $\lambda\geq0$ is a parameter and $\|\Phi\|^2:=-\sfrac12\Tr\Phi^2$.  A monopole is a finite-energy solution of its Euler-Lagrange equations satisfying the boundary condition
\[ \|\Phi\| \to 1 \quad\mbox{as }r\to\infty. \]
The asymptotic scalar field of a monopole defines a map from the 2-sphere at infinity to the unit sphere in $\mathfrak{su}(2)$, and the topological charge of the monopole is the winding number $N$ of this map.

The non-vanishing asymptotic value for $\|\Phi\|$ breaks the gauge symmetry from SU(2) to U(1).  Motivated by this, 't Hooft proposed \cite{tHooft74} the following definition of the asymptotic abelian magnetic field of a monopole:
\begin{equation}
\label{tHooft b}
b_i':=\frac14\epsilon_{ijk}\frac{\Tr(F_{jk}\Phi)}{\|\Phi\|} - \frac18\epsilon_{ijk}\frac{\Tr(\Phi D_j\Phi D_k\Phi)}{\|\Phi\|^3}.
\end{equation}
Another commonly accepted definition for the abelian magnetic field is \cite{manton-sutcliffe}
\begin{equation}
\label{asymptotic b}
b_i := \frac14\epsilon_{ijk}\frac{\Tr(F_{jk}\Phi)}{\|\Phi\|}.
\end{equation}
The equations of motion imply that this magnetic field differs from 't Hooft's only by terms which decay exponentially, so these two magnetic fields share the same asymptotic expansion.  Both magnetic fields $b_i$ and $b_i'$ have singularities at points where $\Phi=0$.

The total magnetic charge $g$ of the monopole is defined to be the flux of $b_i$ (or equivalently, of $b_i'$) through the 2-sphere at infinity.  It can be shown that $g=-2\pi N$, so the magnetic charge is topologically quantised.

It is common in the study of monopoles to introduce two twisted Dirac operators with real parameter $s\in(-1,1)$:
\begin{eqnarray*}
 D_s^\dagger &=& \ii\sigma_jD_j + \ii s+\Phi \\
 D_s &=& \ii\sigma_jD_j - \ii s-\Phi.
\end{eqnarray*}
These act on $L^2$-normalisable spinors transforming in the fundamental representation of SU(2).  Let $\psi_1,\psi_2,\ldots,\psi_{n}$ be a basis for the space of solutions to $D_s\psi=0$, let $\chi_1,\chi_2,\ldots\chi_{n'}$ be a basis for the space of solutions to $D_s^\dagger\chi=0$, and suppose that these bases are both orthonormal:
\[ \int \psi_a^\dagger \psi_b\, \dd^3\mathbf{x} = \delta_{ab}\quad\mbox{and}\quad \int \chi_c^\dagger \chi_d\, \dd^3\mathbf{x} = \delta_{cd}. \]
We propose
\begin{equation}
\label{charge density}
\mu_s(\mathbf{x}) = 2\pi\left(-\sum_{a=1}^{n}\psi_a^\dagger \psi_a + \sum_{c=1}^{n'} \chi_a^\dagger\chi_a\right)
\end{equation}
as a definition of the magnetic charge density of a monopole.  Note that this does not depend on the choice of orthonormal bases $\psi_a$ and $\chi_a$.  This density is not unique, as it depends on the parameter $s\in (-1,1)$.  It will be demonstrated below that, although the densities $\mu_s$ may differ, they induce the same asymptotic magnetic field, with the consequence that all values of $s\in(-1,1)$ yield equally viable charge densities $\mu_s$. However, if a unique charge density was required then $\mu_0$ seems the most natural choice.  Note that all of the densities $\mu_s$ decay exponentially in $r$.

The first requirement of any putative magnetic charge density is that its integral should equal the total magnetic charge $-2\pi N$ as viewed from infinity.  Our proposed density meets this requirement: the normalisation conditions above imply that
\[ \int_{\RR^3}\mu_s\,\dd^3x = 2\pi (n'-n), \]
and an index theorem guarantees that $n-n'=\dim\ker D_s-\dim\ker D_s^\dagger=N$.  Thus the topological nature of the magnetic charge is made manifest through the index theorem.

A more sophisticated requirement of a magnetic charge density is that the moments of the density agree with the multipole expansion of the magnetic field \eqref{asymptotic b} in the manner described in the previous section.  The main result of this paper is that our proposed density meets this requirement, at least in the case of BPS monopoles.

\section{Moments and the Nahm transform}
\label{sec4}

BPS monopoles with $N\geq0$ are solutions of the first order equation
\begin{equation}
\label{bog}
D_i\Phi = \frac12\epsilon_{ijk}F_{jk},
\end{equation}
and the boundary condition
\[ \|\Phi\| \sim 1-\frac{N}{2r}\quad\mbox{as }r\to\infty. \]
They solve the second order Euler-Lagrange equations for the Yang-Mills-Higgs energy in the limiting case where $\lambda=0$.

The BPS equation guarantees that
\begin{align*}
D_s D_s^\dagger 
&= D_iD_i + (\ii s+\Phi)^2.
\end{align*}
This operator is negative and therefore $D_s^\dagger$ has no zero-modes $\chi_a$.  Therefore our definition of the magnetic charge density reduces in the case of BPS monopoles to
\begin{equation}
\label{density}
\mu_s = -2\pi \sum_{a=1}^{N}\psi_a^\dagger \psi_a.
\end{equation}

BPS monopoles can be completely constructed through the formalism of the Nahm transform \cite{manton-sutcliffe,hitchin83,cg}.  This transform associates to any monopole the matrix-valued functions
\begin{align*}
(T_j(s))_{ab} &= -\ii\int x_j\psi_a^\dagger(\mathbf{x};s) \psi_b(\mathbf{x};s)\, \dd^3\mathbf{x} \\
(T_0(s))_{ab} &= +\int \psi_a^\dagger(\mathbf{x};s) \frac{\pa}{\pa s}\psi_b(\mathbf{x};s)\, \dd^3\mathbf{x}.
\end{align*}
It is a non-trivial but well-known result that these matrices solve the Nahm equation,
\[ \frac{\dd T_i}{\dd s}+[T_0,T_i] = \frac12\epsilon_{ijk}[T_j,T_k]\quad\mbox{for }i=1,2,3, \]
and certain boundary conditions (whose precise form does not concern us here).

The Nahm equation is equivalent to a Lax equation
\[ \frac{\dd}{\dd s} T(\zeta) + [T_+(\zeta),T(\zeta)] = 0, \]
in which
\begin{align*}
T(\zeta) &= \frac12(T_1+\ii T_2) + \zeta T_3 - \frac12(T_1-\ii T_2)\zeta^2 \\
T_+(\zeta) &= T_0-\ii T_3 + \ii (T_1-\ii T_2) \zeta.
\end{align*}
It follows that the quantities $\Tr\big( (\ii T(\zeta))^\ell \big)$ are independent of $s$, for $\ell=0,1,2,\ldots$.  These conserved quantities are in fact equal to the moments of the density \eqref{density}, as the following proposition shows:
\begin{proposition}
\label{prop:moments charges}
The moments of the charge density \eqref{charge density} of a BPS monopole and the conserved charges of its associated Nahm data satisfy
\begin{equation}
\label{moments charges}
\int_{\RR^3} \mu_s(\mathbf{x})\,x(\zeta)^\ell\,\dd^3x = -2\pi\Tr\big( (\ii T(\zeta))^\ell \big) .
\end{equation}
\end{proposition}
Before presenting the proof we note that this implies that the moments of the charge distribution $\mu_s$ are independent of $s\in(-1,1)$.  This is why we believe all of the densities $\mu_s$ are equally viable candidates for a magnetic charge density.  For future reference, we denote these moments by
\[ M_\ell(\zeta) = \int_{\RR^3} \mu_s(\mathbf{x})\,x(\zeta)^\ell\,\dd^3x. \]
\begin{proof}
The identity \eqref{moments charges} will be proved from standard identities for Green's functions.  Let $G(\mathbf{x},\mathbf{x}';s)$ be the Green's function for $D_s D_s^\dagger$, i.e. the $2\times 2$ matrix-valued function which solves
\[ (-D_iD_i - (\ii s+\Phi)^2)G(\mathbf{x},\mathbf{x}';s) = \delta(\mathbf{x}-\mathbf{x'}). \]
For convenience we introduce the notation
\[ G(\mathbf{x},\mathbf{x}';s) \overleftarrow{D_s'} = \ii\sigma_j\left(-\frac{\pa G(\mathbf{x},\mathbf{x}';s)}{\pa x_j'} + G(\mathbf{x},\mathbf{x}';s) A_j(\mathbf{x}')\right) -G(\mathbf{x},\mathbf{x}';s)(\ii s+\Phi(\mathbf{x}')). \]
The following identity is well-known (cf.\ equation (4.35), (4.36) in \cite{cg}):
\begin{equation*}
\label{greens function identity}
 \sum_{a=1}^N\psi_a(\mathbf{x};s)\psi_a^\dagger(\mathbf{x}';s) = \delta_3(\mathbf{x}-\mathbf{x}') - D_s^\dagger G(\mathbf{x},\mathbf{x}';z) \overleftarrow{D_s'}.
\end{equation*}

We will use induction and this identity to prove the statement
\begin{equation}
 \label{2}
 ((\ii T(\zeta))^\ell)_{ab} = \int \psi_a^\dagger(\mathbf{x};s)\psi_b(\mathbf{x};s) x(\zeta)^\ell\,\dd^3\mathbf{x}\quad\forall l\in\ZZ,\,l\geq 0,
\end{equation}
from which our main result \eqref{moments charges} follows.

The case $\ell=0$ of \eqref{2} follows directly from the normalisation of the fermion zero modes.

Suppose then that (\ref{2}) holds in the case $\ell=m$ for some $m\in\mathbb{Z}$.  We will show that it must also hold in the case $\ell=m+1$.  Appealing to the Greens' function identity \eqref{greens function identity} and integrating by parts yields:
\begin{align*}
((&\ii T(\zeta))^{m+1})_{ab} \\
&= \int x(\zeta)^m\psi_a^\dagger(\mathbf{x})\psi_c(\mathbf{x})\psi_c^\dagger(\mathbf{x}')\psi_b(\mathbf{x}')  x'(\zeta)\,\dd^3\mathbf{x}\,\dd^3\mathbf{x}' \\
&= \int \psi_a^\dagger(\mathbf{x}) \left( \delta_3(\mathbf{x}-\mathbf{x}') - D_s^\dagger G_s(\mathbf{x},\mathbf{x}') \overleftarrow{D'_s}\right) \psi_b(\mathbf{x}') x(\zeta)^mx'(\zeta)\,\dd^3\mathbf{x}\,\dd^3\mathbf{x}' \\
&= \int \psi_a^\dagger(\mathbf{x})\psi_b(\mathbf{x}) x(\zeta)^{m+1}\,\dd^3\mathbf{x} \\&\qquad 
- \int \big(D_s\overline{x(\zeta)}^m\psi_a(\mathbf{x})\big)^\dagger\big(D'_sx'(\zeta)\psi_b(\mathbf{x}')\big) \,\dd^3\mathbf{x}\,\dd^3\mathbf{x}'.
\end{align*}
Observe that $[D_s,x(\zeta)]=\sigma(\zeta)=\left[D_s,\overline{x(\zeta)}\right]^\dagger$, where
\[ \sigma(\zeta):=\frac{1}{2}(\sigma_2+\ii\sigma_3)+\sigma_1\zeta-\frac{1}{2}(\sigma_2-\ii\sigma_3)\zeta^2. \]
Since in addition $D_s\psi_a=0$, the unwanted second term on the right equates to
\begin{multline*}
\int \big(D_s\overline{x(\zeta)}^m\psi_a(\mathbf{x})\big)^\dagger\big(D'_sx'(\zeta)\psi_b(\mathbf{x}')\big) \,\dd^3\mathbf{x}\,\dd^3\mathbf{x}' \\ 
= m \int  \psi_a(\mathbf{x})^\dagger \sigma(\zeta)^2x(\zeta)^{m-1} \psi_b(\mathbf{x}') \,\dd^3\mathbf{x}\,\dd^3\mathbf{x}'.
\end{multline*}
This expression vanishes, because $\sigma(\zeta)^2=0$.  Therefore the identity \eqref{2} holds in the case $\ell=m+1$, and for all $\ell\geq 0$ by the principle of mathematical induction.
\end{proof}

\section{Higgs field asymptotics}
\label{sec5}

In the case of BPS monopoles the norm of the scalar field $\Phi$ provides a scalar potential for the asymptotic magnetic field \eqref{asymptotic b}: it is easily shown using \eqref{bog} that $b_i=-\pa_i \|\Phi\|$.

Hurturbise has derived \cite{hurtubise} an expression for the asymptotic behaviour of $\|\Phi\|$ in terms of spectral curves.  We recall that the spectral curve of a BPS monopole is the vanishing set of the polynomial
\[ g(\eta,\zeta)=\det(\eta-\ii T(\zeta)). \]
We note that, like the polynomials $\Tr((\ii T(\zeta))^\ell)$, this polynomial $g$ is independent of $s$.
\begin{definition}
The \emph{tail} of a BPS monopole is a real function on the complement of a compact subset of $\RR^3$, defined by the following contour integral:
\begin{equation}
\label{hurtubise}
\curlyv:=-\frac{1}{4\pi\ii} \oint_\Gamma \frac{\pa_\eta g(\eta,\zeta)}{g(\eta,\zeta)} \Big|_{\eta=x(\zeta)}\dd\zeta.
\end{equation}
For sufficiently large $r$, half of the poles of the integrand cluster near the point $\zeta_+=-(x_2+\ii x_3)/(r+x_1)$ on the Riemann sphere corresponding to $\mathbf{x}/r$, and half near its antipode $\zeta_-=-1/\bar{\zeta}_+$.  The contour $\Gamma$ encloses the former and not the latter.  The domain of $\curlyv$ is chosen such that none of the poles move from one cluster to another as $\mathbf{x}$ moves through the domain
\end{definition}
\begin{theorem}[Hurtubise \cite{hurtubise}]
\label{thm:hurtubise}
The norm of the Higgs field and tail of a BPS monopole satisfy
\[ \|\Phi\|=1+\curlyv \]
up to exponentially decaying terms.
\end{theorem}

Hurtubise' result can be used to prove:
\begin{theorem}
\label{thm:main}
Let $(A,\Phi)$ be a BPS monopole, let $s\in(-1,1)$, and let $\mu_s$ be the charge density defined in equation $\eqref{charge density}$.  Let $\phi_s$ be a solution to $\triangle\phi_s=-\mu_s$ and suppose that it admits an asymptotic expansion in powers of $1/r$.  Then the asymptotic expansions of $\phi_s$ and $\|\Phi\|-1$ agree to all orders.
\end{theorem}
\noindent
The physical interpretation of this theorem is that the multipole expansion of the magnetic field induced by $\mu_s$ agrees with that of the magnetic field $b_i$ defined in \eqref{asymptotic b}.

\begin{proof}
The tail is automatically harmonic almost everywhere, as follows from the formalism of the Penrose transform.  Therefore it admits an asymptotic expansion in powers of $1/r$ of the form
\begin{equation}
\label{v series}
\curlyv = \sum_{\ell=0}^\infty \frac{\curlyv_\ell}{r^{\ell+1}},
\end{equation}
in which $\curlyv_\ell$ are spherical harmonics of weight $\ell$.  By Hurtubise' theorem \ref{thm:hurtubise} the function $\|\Phi\|-1$ admits an expansion in powers of $1/r$ that agrees precisely with this expansion of $\curlyv$.

It is straightforward to derive expressions for these functions $\curlyv_\ell$ from the integral expression \eqref{hurtubise}.  First, note that
\[ \frac{\pa_\eta g(\eta,\zeta)}{g(\eta,\zeta)} = \Tr \big((\eta-\ii T(\zeta))^{-1}\big). \]
Therefore
\begin{align*}
\frac{-1}{4\pi\ii} \oint_\Gamma \frac{\pa_\eta g(\eta,\zeta)}{g(\eta,\zeta)} \Big|_{\eta=x(\zeta)}\dd\zeta
&=  \frac{-1}{4\pi\ii}\oint_\Gamma \Tr \big((x(\zeta)-\ii T(\zeta))^{-1}\big) \dd\zeta \\
&= \sum_{\ell=0}^\infty \frac{-1}{4\pi\ii}\oint_\Gamma \frac{\Tr ((\ii T(\zeta))^\ell) }{x(\zeta)^{\ell+1}}\dd\zeta.
\end{align*}
Since $M_\ell(\zeta)=-2\pi\Tr((\ii T(\zeta))^\ell)$, we conclude that
\begin{equation}
\label{M to v}
\curlyv_\ell = \frac{1}{8\pi^2\ii}\oint_\Gamma \frac{M_\ell(\zeta)}{n(\zeta)^{\ell+1}}\dd\zeta.
\end{equation}

Suppose now that $\phi_s$ admits an expansion of the form $\phi_s=\sum_{\ell=0}^\infty\phi_\ell/r^{\ell+1}$.  By propositions \ref{prop:moments asymptotics} and \ref{prop:moments charges} the coefficients $\phi_\ell$ satisfy
\[ \phi_\ell = \frac{1}{8\pi^2\ii}\oint_\Gamma \frac{M_\ell(\zeta)}{n(\zeta)^{\ell+1}}\dd\zeta. \]
Thus the expansions of $\phi_s$ and $\|\Phi\|-1$ coincide.
\end{proof}

In the next section we present some results concerning the explicit evaluation of the tail function $\curlyv$.  Before doing so, we pause to point out that the Nahm transform provides a natural definition of the potential function for the magnetic field induced by $\mu_s$.  We recall that the Nahm data Green's function $f(s,s';\mathbf{x})$ is the $N\times N$ matrix-valued solution to
\[
-\left( \left(\frac{\dd}{\dd s}+T_0\right)^2 + \left(T_j+\ii x_j\right)^2\right) f(s,s';\mathbf{x}) = \mathrm{Id}_N\delta(s-s').
\]
This Green's function is related to the monopole zero-modes $\psi_a$ by the identity \cite{cg},
\[
2\pi \psi_a^\dagger(\mathbf{x};s)\psi_b(\mathbf{x};s') = ((s-s')^2-\triangle)f_{ab}(s,s';\mathbf{x}).
\]
It follows that
\[
\mu_s = \triangle\Tr f(s,s;\mathbf{x}).
\]
Therefore $\phi_s=-\Tr f(s,s;\mathbf{x})$ is a potential for the magnetic field induced by $\mu_s$.

In \cite{BNvB, nogradi} it was conjectured that the asymptotic expansion for $\Tr f(s,s;\mathbf{x})$ is determined to all orders by the
conserved quantities $\Tr((\ii T(\zeta))^\ell)$ of the Nahm equation.  This conjecture provided the initial motivation
for our investigations.  It can be proved directly from propositions \ref{prop:moments asymptotics} and \ref{prop:moments charges}, as in the proof of theorem \ref{thm:main}.


\section{Evaluating the asymptotic Higgs field}
\label{sec6}

In general reconstructing a monopole from its Nahm data is a difficult problem.  In this section we show that the tail, and hence the asymptotics, of a monopole can straightforwardly be evaluated from its spectral curve or the conserved tensors of the Nahm equation.  Explicit formulae will be presented for examples with Platonic symmetry.

The first step in evaluating the tail is to determine the moments $M_\ell(\zeta)$ from the spectral curve.  Let $g_m(\zeta)$ be the coefficient of $\eta^{N-m}$ in the spectral curve, so that
\[ 
g(\eta,\zeta) = \sum_{m=0}^N g_m(\zeta)\eta^{N-m}.
\]
Note in particular that $g_0=1$.  Newton's identity states that
\[
g_\ell(\zeta)=\frac{1}{2\pi \ell} \sum_{m=1}^\ell g_{\ell-m}(\zeta)M_\ell(\zeta)
\]
Rearranging this yields the formula
\begin{equation}
M_\ell(\zeta)=-\sum_{m=1}^{\ell-1} g_{\ell-m}(\zeta)M_m(\zeta)+ 2\pi\ell g_\ell(\zeta),
\end{equation}
in which it should be understood that $g_m=0$ for $m>N$.  From this formula the polynomials $M_\ell(\zeta)$ can be calculated recursively.

The second step in evaluating the tail is to determine the spherical harmonics $\curlyv_\ell$ from the polynomials $M_\ell$.  In principle this can be achieved by evaluating the contour integral \eqref{M to v}, but in practice it is useful to have explicit formulae in terms of the basis polynomials $\zeta^0,\zeta^1,\ldots,\zeta^{2\ell}$.  The following lemma provides such formulae.

\begin{lemma}
\label{lem:M to v basis}
Let $M_\ell(\zeta)=\sum_{m=-\ell}^\ell M_\ell^m\zeta^{\ell+m}$ and let $\curlyv_\ell$ be the function on $S^2$ obtained from $M_\ell$ by the contour integral \eqref{M to v}.  Let $Y_\ell^m:S^2\to\RR$ by the functions defined by
\[ 
\left(\frac{-n(\zeta)}{\zeta^2}\right)^\ell = \sum_{m=-\ell}^\ell \frac{1}{(-\zeta)^{\ell+m}} Y_\ell^m.
\]
Then
\[ \curlyv_\ell
= \sum_{m=-\ell}^\ell M_\ell^m \frac{(\ell-m)!(\ell+m)!}{(4\pi)(\ell!)^2} Y_\ell^m.
\]
\end{lemma}
\noindent
Note that the functions $Y_\ell^m$ agree up to normalisation and rotation with the standard spherical harmonics.

\begin{proof}
We begin by noting that
\[ 
\left(\frac{-n(\zeta)}{\zeta^2}\right)^\ell =
\exp\left(\frac{L_+}{\zeta}\right)\left(\frac{n_1-\ii n_2}{2}\right).
\]
The case $\ell=1$ of this identity can be verified by direct calculation, and the cases with $\ell>1$ follow because $L_+$ obeys the Leibniz rule.  It follows that
\[ Y_\ell^m = \frac{(-L_+)^{\ell+m}}{(\ell+m)!}\left(\frac{n_1-\ii n_2}{2}\right). \]

To evaluate the contour integral \eqref{M to v}, we note the following identity (which is easily proved):
\begin{equation*}
\zeta^{\ell+m}=\frac{(\ell-m)!}{(2\ell)!}(-J_+^{(\ell)})^{\ell+m}\zeta^0.
\end{equation*}
Here $J_+^{(\ell)}$ is the operator defined in the proof of proposition \ref{prop:moments charges}.  It follows from this identity and equations \eqref{J and L 2} and \eqref{oint 1/n^l} that
\begin{align*}
\frac{1}{8\pi^2\ii}\oint_\Gamma \frac{\zeta^{\ell+m}}{n(\zeta)^{\ell+1}}\dd\zeta
&= \frac{(\ell-m)!}{(2\ell)!}(-L_+)^{\ell+m} \frac{1}{8\pi^2\ii}\oint_\Gamma \frac{1}{n(\zeta)^{\ell+1}}\dd\zeta \\
&= \frac{(\ell-m)!}{(4\pi)2^\ell(\ell!)^2} (-L_+)^{\ell+m}(n_1-\ii n_2)^\ell \\
&= \frac{(\ell-m)!(\ell+m)!}{4\pi(\ell!)^2} Y_\ell^m.
\end{align*}
The result follows.

\end{proof}

\begin{figure}[htb]
\begin{center}
 \includegraphics[width=6cm]{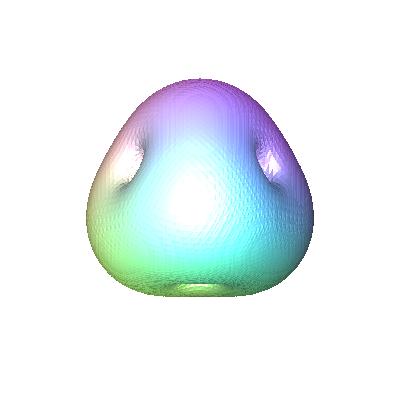}
 \includegraphics[width=6cm]{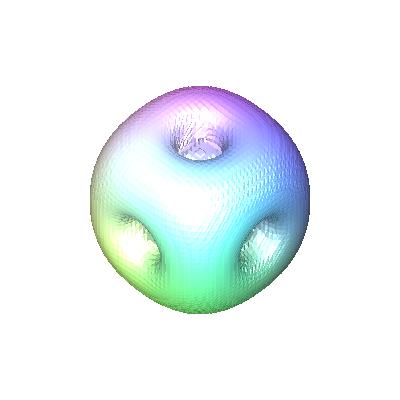}\\
 \includegraphics[width=6cm]{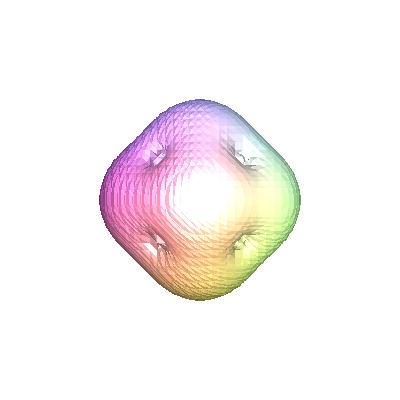}
 \includegraphics[width=6cm]{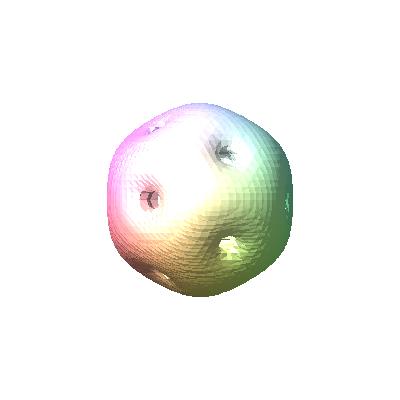}
\end{center}
\label{figure}
\caption{Isosurfaces of $\pa_i\pa_iv$ for the Platonic monopoles with charges 3 (top left), 4 (top right), 5 (bottom left), and 7 (bottom right).}
\end{figure}

Lemma \ref{lem:M to v basis} provides a means to evaluate $\curlyv_\ell$ from $M_\ell$ which is
easily implemented in standard algebraic software packages.  Below we present series expansions of $\curlyv$ for
examples of monopoles with Platonic symmetry, along with closed form expressions along certain symmetry axes.  In figure
\ref{figure} we display isosurfaces for $\pa_i\curlyv\pa_i\curlyv$, which approximates the energy density of the monopole.  These are created using series expansions for $\curlyv$ up to $\ell=12$ (removing the last non-zero term from the series expansion did not alter the pictures).  They closely resemble the pictures published in \cite{houghton-sutcliffe96a,houghton-sutcliffe96b}, so it seems that much of the structure of a monopole is captured by its abelian tail.

\subsection{The tetrahedral 3-monopole}

There is a unique monopole with topological charge $N=3$ and tetrahedral symmetry \cite{hmm, houghton-sutcliffe96a}.  Its spectral curve is defined by the polynomial
\begin{align*}
g(\eta,\zeta) &= \eta^3 - \ii C_3 \zeta(\zeta^4-1),\quad\mbox{where} \\
C_3 &= \frac{2\pi^6}{3^{9/2}\Gamma(2/3)^9}.
\end{align*}
The first few terms in the expansion \eqref{v series} for $\curlyv$ are easily evaluated.  In order to write them down in a concise manner, we make use of the fact that all tetrahedrally-symmetric polynomial functions on the two-sphere can be written in terms of the three polynomials $t_2=n_1^2+n_2^2+n_3^2=1$, $t_3=n_1n_2n_3$ and $t_4=n_1^4+n_2^4+n_3^4$.  In these terms, we find that
\[
\curlyv = -\frac{3}{2r} + \frac{15 C_3}{r^4} t_3 + \frac{3C_3^2}{4r^7}(17t_2^3-21t_2t_4-462t_3^2)+\ldots
\]
Note that in general the terms $\curlyv_\ell$ vanish except when $\ell=0\mod 3$.  This property does not follow from
symmetry considerations alone: for example, there is a tetrahedrally-symmetric spherical harmonic with $\ell=4$, but
this spherical harmonic does not appear in the series expansion for $\curlyv$.

It seems difficult to sum the series expansion for $\curlyv$ in general.  However, we have been able to find a closed form expression in some special cases, by working directly from the contour integral expression \eqref{hurtubise}.  Restricting eq.\ \eqref{hurtubise} to the $x_1$-axis $(x_1,x_2,x_3)=(t,0,0)$ yields the expression
\[
\curlyv = -\frac{1}{4\pi\ii}\oint_\Gamma \frac{3t^2\zeta\dd\zeta}{-\ii C_3\zeta^4+t^3\zeta^2+\ii C_3}.
\]
By change of variables $w=\zeta^2$, the contour integral becomes
\[
\curlyv = -\frac{1}{4\pi\ii}\oint_{\Gamma'} \frac{3t^2\dd w/2}{-\ii C_3w^2+t^3w+\ii C_3}.
\]
The denominator of the integrand has two roots, but only the root $w=(t^3-\sqrt{t^6-C^2})/2\ii C$ lies inside the contour $\Gamma'$.  The contour $\Gamma'$ circles this pole twice because $w=\zeta^2$.  Therefore the integral evaluates to
\[
\curlyv = -\frac{3t^2}{2\sqrt{t^6-4C_3^2}}.
\]
It is straightforward to check that the asymptotic expansion of this function agrees with the restriction of the series
expansion for $\curlyv$ to the $x_1$-axis.

The reason why a closed form expression can be obtained on the $x_1$-axis is that this line has a high degree of symmetry.  The symmetry group of the monopole fixes a tetrahedron, and the $x_1$-axis passes through opposite edges of this tetrahedron.  Rotations through $\pi$ about the $x_1$-axis fix both the axis and the monopole.  This rotational symmetry allowed a simplification of the contour integral, so that the roots of the denominator could be easily found.

There is another line with a high degree of symmetry, namely that passing through a vertex and the centre of the
opposing face of the tetrahedron.  This line has equation $(x_1,x_2,x_3)=(t,t,t)/\sqrt{3}$.  A closed form expression
for $\curlyv$ can also be obtained along this line.

To obtain the expression for $\curlyv$ it is convenient to first rotate the monopole so that the desired line is again the $x_1$-axis.  This is accomplished by making a M\"obius transform of the spectral curve of the form
\[ g(\eta,\zeta)\mapsto (-\bar{b}\zeta+\bar{a})^{2N}g\big(\eta/(-\bar{b}\zeta+a),(a\zeta+b)/(-\bar{b}\zeta+a)\big)
\]
with $a=e^{-\pi\ii/8}\sqrt{(1+1/\sqrt{3})/2}$ and $b=-e^{-\pi\ii/8}\sqrt{(1-1/\sqrt{3})/2}$.  After making this M\"obius transformation the spectral curve becomes
\[
g(\eta,\zeta)=\eta^3-\frac{C_3}{\sqrt{27}}(\sqrt{2}\zeta^6+10\zeta^3-\sqrt{2}).
\]
With $g$ in this form and $(x_1,x_2,x_3)=(t,0,0)$ the contour integral \eqref{hurtubise} becomes
\[
\curlyv=-\frac{1}{4\pi\ii}\oint \frac{3t^2\zeta^2\dd\zeta}{-\sqrt{2/27}C_3\zeta^6+(10\sqrt{27}C_3+t^3)\zeta^3+\sqrt{2/27}C_3}.
\]
This integral is simplified by the substitution $w=\zeta^3$, and evaluates to
\[
\curlyv = -\frac{3t^2}{2\sqrt{t^6+(20/\sqrt{27})C_3t^3+4C_3^2}}.
\]
We have checked that the expansion of this expression in powers of $1/t$ agrees with the restriction of the series expansion for $\curlyv$ to the line $(x_1,x_2,x_3)=(t,t,t)/\sqrt{3}$.

\subsection{The cubic 4-monopole}

The unique monopole with topological charge 4 and cubic symmetry has spectral curve \cite{houghton-sutcliffe96a}
\begin{align*}
g(\eta,\zeta) &= \eta^4 + C_4 (\zeta^8+14\zeta^4+1),\quad\mbox{where} \\
C_4 &= \frac{3\pi^6}{2^8\Gamma(3/4)^8}.
\end{align*}
The first few terms in its series expansion for $\curlyv$ are easily calculated, and can be written in terms of the octahedrally-symmetric polynomials $o_2=t_2$, $o_4=t_4$ and $o_6=t_3^2$ as
\[
\curlyv = -\frac{4}{2r} + \frac{14C_4}{r^5}(5o_4-3o_2^2) + \frac{99C_4^2}{r^9}(208o_6o_2+94o_4o_2^2-65o_4^2-33o_2^4)+\ldots.
\]
We have been able to evaluate $\curlyv$ in closed form along lines which pass through opposing vertices and opposing faces of the cube fixed by the symmetry group:
\begin{align*}
\curlyv(t,0,0) &= -\frac{2t^3}{\sqrt{t^8+28C_4t^4+192C_4^2}} \\
\curlyv\left(\frac{t}{\sqrt{3}},\frac{t}{\sqrt{3}},\frac{t}{\sqrt{3}}\right) &= -\frac{2t^3}{\sqrt{t^8-56C_4t^4/3+144C_4^2}}.
\end{align*}
The series expansions of these functions in powers of $1/t$ agree with the general series expansion quoted above.  Both of these expressions are obtained using a similar method to the one used for the 3-monopole.  For the second, it is convenient to first make a M\"obius transformation with  $a=e^{-\pi\ii/8}\sqrt{(1+1/\sqrt{3})/2}$ and $b=e^{-\pi\ii/8}\sqrt{(1-1/\sqrt{3})/2}$, after which the spectral curve has the form
\[ g(\eta,\zeta) = \eta^4-\frac{4C_4}{3}(2\sqrt{2}\zeta^6+7\zeta^3-2\sqrt{2})\zeta \]
and the desired line has moved to $(x_1,x_2,x_3)=(t,0,0)$.

\subsection{The octahedral 5-monopole}

The unique monopole with topological charge 5 and cubic symmetry has spectral curve
\begin{align*}
g(\eta,\zeta) &= \eta^5 - C_5 (\zeta^8+14\zeta^4+1)\eta,\quad\mbox{where} \\
C_5 &= \frac{3\pi^6}{2^6\Gamma(3/4)^8}.
\end{align*}
Note that this differs from the result quoted in \cite{houghton-sutcliffe96b} in that the coefficient of the polynomial in $\zeta$ is $-C_5$ rather than $C_5$ -- in attempting to reproduce the calculation of \cite{houghton-sutcliffe96b} we discovered a sign error.

The first few terms in the series expansion for $\curlyv$ are
\[
\curlyv = -\frac{5}{2r} + \frac{14C_5}{r^5}(3o_2^2-5o_4) + \frac{99C_5^2}{r^9}(208o_6o_2+94o_4o_2^2-65o_4^2-33o_2^4)+\ldots.
\]
We have been able to evaluate $\curlyv$ in closed form along lines which pass through opposing vertices and opposing faces of the octahedron fixed by the symmetry group:
\begin{align*}
\curlyv(t,0,0) &= -\frac{1}{2t}-\frac{2t^3}{\sqrt{t^8-28C_5t^4+192C_5^2}} \\
\curlyv\left(\frac{t}{\sqrt{3}},\frac{t}{\sqrt{3}},\frac{t}{\sqrt{3}}\right) &= -\frac{1}{2t}-\frac{2t^3}{\sqrt{t^8+56C_5t^4/3+144C_5^2}}.
\end{align*}
The series expansions of these functions in powers of $1/t$ agree with the general series expansion quoted above.  Both of these expressions are obtained using a similar method to the one used for the 3-monopole.  For the second, it is convenient to first make a M\"obius transformation as for the 4-monopole, after which the spectral curve has the form
\[ g(\eta,\zeta) = \eta^5+\frac{4C_5}{3}(2\sqrt{2}\zeta^6+7\zeta^3-2\sqrt{2})\zeta\eta. \]

\subsection{The dodecahedral 7-monopole}

The unique charge 7 monopole with dodecahedral symmetry has spectral curve,
\begin{align*}
g(\eta,\zeta) &= \eta^7 - C_7(\zeta^{11}-11\zeta^6-\zeta)\eta,\quad\mbox{where} \\
C_7 &= \frac{16\pi^{12}}{729\Gamma(2/3)^{18}}.
\end{align*}
The first few terms	in the series expansion for $\curlyv$ are
\begin{align*}
\curlyv &= -\frac{7}{2r} + \frac{33 C_7 i_6}{16r^7} + \ldots ,\quad\mbox{where} \\
i_6 &= 16x_1^6-120x_1^4(x_2^2+x_3^2)+90x_1^2(x_2^2+x_3^2)^2 \\ & \quad -42x_1(x_2^5-10x_2^3x_3^2+5x_2x_3^4)-5(x_2^2+x_3^2)^3.
\end{align*}
The tail function $\curlyv$ has the following closed form expression along the line $(x_1,x_2,x_3)=(t,0,0)$:
\[ \curlyv = -\frac{1}{2t} - \frac{3t^6}{\sqrt{t^{12}+22C_7t^6+125C_7^2}}. \]
The series expansion of this function in $1/t$ agrees with the general series expansion quoted above.

The line on which we have evaluated $\curlyv$ passes through the centres of opposite faces of the dodecahedron fixed by
the symmetry group.  One might ask whether it is also possible to evaluate $\curlyv$ on a line passing through opposite vertices.  In order to do so using the method above one would first need to move this line to the $x_1$-axis by M\"obius transformation and then to factor the denominator in the contour integral.  On symmetry grounds this denominator is a quartic polynomial in $\zeta^3$, which could in principle be factorised, but we have not attempted to do so.

\section{Conclusion}
\label{sec7}

In this paper we have proposed a novel definition \eqref{charge density} for the magnetic charge density of a monopole with gauge group SU(2).  This definition differs from standard definitions in that it is smooth and non-singular.  We have shown that, in the case of BPS monopoles, the abelian magnetic field which it induces agrees with the standard definitions \eqref{tHooft b} and \eqref{asymptotic b} to all orders in the multipole expansion.

This result can straightforwardly be extended to the case of SU($n$) monopoles.  SU($n$) monopoles with maximal symmetry-breaking have $n-1$ asymptotic abelian magnetic fields.  Hurtubise and Murray have proved in \cite{hurtubise-murray} a result which relates the asymptotics of these magnetic fields to the spectral curves and which generalises the result of Hurtubise employed in this paper.  The definition \eqref{charge density} of the magnetic charge densities $\mu_s$ generalises directly to the SU($n$), as do our arguments relating these to the Nahm data conserved quantities and the asymptotic magnetic fields.  The chief novelty is that different values of $s$ must be chosen to source the different magnetic fields.  It may be possible to generalise our result to apply to SU($n$) calorons also, however, the analogue of Hurtubise' result for calorons is not currently known.

The tail function, which describes the asymptotics of the norm of the Higgs field, played a key role in our analysis.  While the fields of a monopole are in general difficult to construct in explicit form, the tail function is easily evaluated as a series given knowledge of the monopole's spectral curve.  Moreover, we have exhibited a number of examples in which this tail function can be evaluated in closed form when restricted to a well-chosen line with a high degree of symmetry.  The problem of reconstructing a monopole field explicitly from its Nahm data remains an important problem, and our observation suggests that more progress might be made if attention is restricted to these symmetric lines.

Our results have some consequence for the analysis of magnetic bags.  One of us argued in \cite{harland} that the Nahm transform for monopoles should converge in a certain large $N$ limit to what was called the $\mathfrak{u}(\infty)$ Nahm transform.  It seems that Hurtubise' result, together with our observation relating the tail function to the Nahm data conserved charges, provides a more direct proof of this result.

A final comment concerns other systems that support BPS topological solitons.  Analogues to the charge density $\mu_s$ could conceivably written down for instantons, vortices, and other solitons.  It would be interesting to investigate whether such densities, like the density for monopoles, carry any physical significance.\bigskip
\\
\textbf{Acknowledgements.}  We are grateful to Falk Bruckmann for providing the stimulus for this project.  We also thank Nick Manton and Bernd Schroers for helpful suggestions.

\end{document}